# Packet Score Based Network Security and Traffic Optimization

S.Karthik, I MTech-CEN, Amrita University
K.Saravanan, Lecturer CSE, Erode Sengunthar Engineering College

*Abstract*—One of the critical threat to internet security is Distributed Denial of Service (DDoS). This paper by the introduction of automated online attack classification and attack packet discarding helps to resolve the network security issue by certain level. The incoming packets are assigned scores based on the priority associated with the attributes and on comparison with probability distribution of arriving packets on per packet basis.

## I. INTRODUCTION

THE major concern of network security being Distributed Denial-of-Service (DDoS), serves to be the root for development in modern day security over networks. The same being the reason for, Packet Based Network Security and Traffic Optimization.

Considering attack traffic filtering, the most vital part of network problems research area the attacks can be classified as

Source-initiated: Source sites are responsible for guaranteeing that outgoing packets are attack-free. Examples include network ingress filters [8]. Disabling ICMP or removing unused services to prevent the computers from becoming attack agents, or filtering unusual traffic from the source [24]. However, the viability of these approaches hinges on voluntary cooperation among a majority of ingress network administrators Internet-wide, making these approaches rather impractical given the scale and uncontrollability of the Internet.

Path-based: In this approach, only the packets following the correct paths are allowed [15]. Any packet with a wrong source IP for a particular router port is considered a spoofed packet and dropped, which eliminates up to 88 percent of the spoofed packets [4], [28]. In another approach [11], if the number of traveled hops is wrong for a source IP, the packet is dropped, thereby eliminating up to 90 percent of the spoofed packets. These approaches are considered practical, but they have a somewhat high probability of false negatives, i.e., falsely accepting attack packets. Apparently, when packets use unspoofed addresses, which are an emerging trend, none of these approaches works.

Victim-initiated: The victim can initiate counter-measures to reduce incoming traffic. For example, in the pushback scheme [10], the victim starts reducing excessive incoming traffic and requests the upstream routers to perform rate reduction as well. There are other methods based on an overlay network [14], packet marking [18], [31], TCP flow filtering [17], [33] and statistical processing [19], [20], etc. Although victim-initiated protections are more desirable, some methods require changes in Internet protocols or are too expensive to implement.

The Packet Score scheme has been proposed recently by the authors of this paper [3], [19]. One of the key concepts in Packet Score is the notion of "Conditional Legitimate Probability" (CLP) based on Bayesian theorem. CLP indicates the likelihood of a packet being legitimate by comparing its attribute values with the values in the baseline profile. Packets are selectively discarded by comparing the CLP of each packet with a dynamic thresh- old. This paper explains the concept of calculating the packet score based in the attribute values and arriving packet probability distribution.

In this paper, we extend the basic concept to a practical real-time packet filtering scheme using elaborate processes. In this paper, we describe the Packet Score operations for single-point protection, but the fundamental concept can be extended to a distributed implementation for core-routers.

The rest of this paper is organized as follows: In Section 2, we describe the concept of Conditional Legitimate Probability (CLP) and Arriving Packet Probability distribution. In Section 3, we focus on the determining normal traffic characteristics. In Section 4, score assignment to packets, selective discarding, and overload control are described. In Section 5, an integrated process combining Sections 2, 3, and 4 is described. The paper concludes in Section 6 with the direction of future investigation

## II. CALCULATING CONDITIONAL LEGITIMATE PROBABILITY AND THE ARRIVING PACKET PROBABILITY DISTRIBUTION

The most crucial part of identifying a DDoS attach is by classifying the attacking packets and the ones that are legitimate as per the source paper. The concept of Conditional Legitimate Probability (CLP) h e l p s  i n  identifying attack packets probabilistically. CLP is produced by comparing traffic characteristics during t h e  attack  with previously measured, l e g i t i m a t e traffic characteristics. The viability of this approach is based on the premise that there are some traffic characteristics that are inherently stable during normal network operations of a target network.

The concept of calculating the Probability distribution of Arriving Packet helps to estimate the nature of the arriving packet at a certain instance of time or for a given source IP. Comparisons leading to an observation of arriving packet nature for a better classification among the classes.

We named this scheme Packet Score because CLP can be viewed as a score which estimates the legitimacy of a suspicious packet. We will use the terms CLP and score interchangeably. By taking a score-based filtering approach, the prioritization of different types of suspicious packets is possible. The ability to prioritize becomes even more important when a full characterization of attack packets is not feasible. By dynamically adjusting the cutoff score according to the available traffic capacity of the victim, our approach allows the victim system to accept more potentially legitimate traffic. In contrast, once a rule-based filtering scheme is configured to discard specific types of packets, it does so regardless of the victim network's available capacity.

To formalize the concept of CLP, we consider all the packets destined for a DDoS attack target. Each packet would carry a set of discrete-value attributes A; B;C; . . . . For example, A might be the protocol type, B might be the packet size, C might be the TTL values, etc. We defined {a1; a2; a3; . . .} as the possible values for attribute A, {b1; b2; b3; . . .} as the possible values for attribute B, and so on. During an attack, there are Nn legitimate packets and Na attack packets arriving in T seconds, totaling Nm.

$$Nm = Nn + Na$$

(m for measured; n for normal; and a for attack)

Pn, the ratio or the probability of attribute values among the legitimate packets, is defined as follows:

$$\sum P_n (A = a_i) = 1$$

The Conditional Legitimate Probability (CLP) is defined as the probability of a packet being legitimate given its attributes:

CLP(packet p) = P(packet p is legitimate | p's attribute

$$A=a_p, attribute\ B = b_p...).$$

According to Bayes' Theorem,

$$CLP(p) = \frac{N_n x P_n (A = a_p) x P_n (B = b_p) x...}{N_m x P_m (A = a_p) x P_m (B = b_p) x...}$$

:

While we leave the investigation on the independence assumption as a future work, it seems to work in practice because the CLP (p) is still a good metric for packet prioritization. A large portion of DDoS attack packets get lower CLPs because Pm becomes larger than Pn for the dominant attribute values in the attack. As long as we can assign lower scores to the majority of attack packets, the assumption of independence is not essential to Packet Score operation. The packet arrival probability helps to discard the attacking packets based on the history store in the database.

### III.   DETERMINING THE NORMAL TRAFFIC

The possibilities to determine exactly the number of legitimate packets while on attack period is quite less. Instead they are determined while at normal traffic operations.

A normal traffic profile consists of single and joint distributions of various packet attributes from IP headers are:

1. Packet size.
2. Time to Live
3. Protocol type values and
4. Source IP prefixes
   Those from TCP headers are:
5. TCP flag patterns and
6. Server port number

The principle of Packet Score is to punish the traffic whose attribute value ratio is higher than in profile. Therefore, to accommodate an occasional surge of particular attribute values in legitimate traffic, the highest ratio among the periodic ratios is selected. This strategy has little impact on blocking attack traffic while giving the legitimate traffic a safety margin.

Packet Score depends on the stability of the traffic profile for estimating Pn. It has been known that for a given subnet, there is a distinct traffic pattern in terms of packet attribute value distribution for a given time and/or given day [12], [21], [23]. In general, the nominal traffic profile is believed to be a function of time which exhibits periodic, time-of-day, and day-of-the-week variations as well as long-term trend changes.

### IV.   SCORE ASSIGNMENT

Scoring a packet is equivalent to looking up the scorebooks, e.g., the TTL scorebook, the packet size scorebook, the protocol type scorebook, etc. After looking up the multiple scorebooks, we add up the matching CLP entries in a log-version scorebook. This is generally faster than multiplying the matching entries in a regular scorebook. The small speed improvement from converting a multiplication operation into an addition operation is particularly useful because every single packet must be scored in real-time. This speed improvement becomes more beneficial as the number of scorebooks increases. On the other hand, generating a log-version scorebook may take longer than a regular scorebook generation. However, the scorebook is generated only once at the end of each period and it is not necessary to observe every packet for scorebook generation; thus, some processing delay can be allowed

### 1. Decoupling the Profile Update and Scoring

According to (3), the current packet attribute distributions ($P_m$) have to be updated constantly whenever a packet arrives. To make wire-speed per-packet score computation possible, we decoupled the updating of packet attribute distribution from that of score computation to allow them to be conducted in parallel, but at different time periods. To be more specific, a frozen set of recent profiles at time period $T_1$ is used to generate a set of scorebooks which is used to score the packets arriving at the next time period, $T_2$. Packets arriving at $T_2$ also generate a new profile and scorebook to be used for time period $T_3$. The time-scale of period $T_i$ is longer than the per-packet arrival time-scale. It can be configured to a fixed length or until the detection of a significant change in the measured traffic profile.

This decoupling introduces a small challenge in catching up with attack profile change. In most cases, the traffic characteristics in adjacent periods are very similar, but during a rapidly changing attack, this assumption may be inaccurate. As a result, the scorebook at $T_i$ does not represent the true scorebook at $T_{ip1}$, and the Packet Score performance degrades. This can be resolved easily by reducing the time-scale of $T_i$ or by using a packet number based period instead of a time-based one.

More over the probability distribution of arriving packet is stored as a training set for future classification of arriving packet based on some devised constraints.

### 2. Packet Discarding

Once the score is computed for a packet, selective packet discarding, and overload control can be performed using the score as the differentiating metric. Since an exact prioritization would require offline, multiple-pass operations, e.g., sorting and packet buffering, we take the following alternative approach. First, we maintain the cumulative distribution function (CDF) of the scores of all incoming packets in time period $T_i$. Second, we calculate the cut-off threshold score Thd as follows which is illustrated in Fig. 1.

- Total current incoming traffic at period $T_i = \psi_i$,
- Acceptable traffic at period $T_i = \phi_i$,
- The fraction of traffic permitted to pass =
  $1 - \Phi_i = \phi_i / \psi_i$, and
- The Thd$_{i+1}$ that satisfies CDF (Thd$_{i+1}$) = $\Phi_i$.

Third, we discard the arriving packets in time period $T_{i+1}$ if its score value is below the cut-off threshold Thd$_{i+1}$. At the same time, the packets arriving at $T_{i+1}$ create a new CDF, and a new Thd$_{i+2}$ is calculated for $T_{i+2}$.

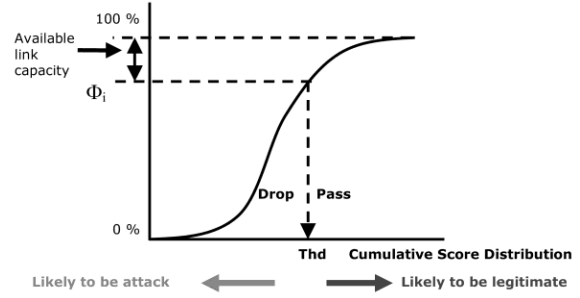

Fig. 1 Selective packet discarding

## V. THE INTEGRATED PROCESS

Fig.2 depicts the integrated operation and the determination of a dynamic discarding threshold. A load-shedding algorithm, such as the one described in [13], is used to determine the amount ($\Phi$) of suspicious traffic arriving that needs to be discarded in order to keep the utilization of the victim below a target value. Typical inputs to a load-shedding algorithm include current utilization of the victim, maximum (target) utilization allowed for the victim, and the current aggregated arrival rate of suspicious traffic. Once the required packet discarding percentage ($\Phi$) is determined, the corresponding CLP discarding threshold, Thd, is determined from a recent snapshot of the CDF of the CLP values. The snapshot is updated periodically or upon significant changes in the packet score distribution. The adjustment of the CLP discarding threshold is done on a time-scale which is considerably longer than the packet arrival time-scale. The entire packet Score process can be best performed in a pipelined approach as discussed in Section 4.2 in which time is divided into fixed intervals, and each operation is performed based on the snapshot of the previous period. Specifically, the following three operations are performed in pipeline when a packet arrives:

1. Incoming packet profiling:
   - Packets are observed to update $P_m$.
   - At the end of the period, $P'_n / P_m$ is calculated and scorebooks are generated.
2. Scoring:
   - The packets are scored according to the most recent scorebooks.
   - . At the end of the period, CDF is generated and the cut-off threshold is calculated.
3. Discarding:
   - The packets are scored according to the most recent scorebooks.
   - The packet is discarded if its score is below the cut-off threshold score.

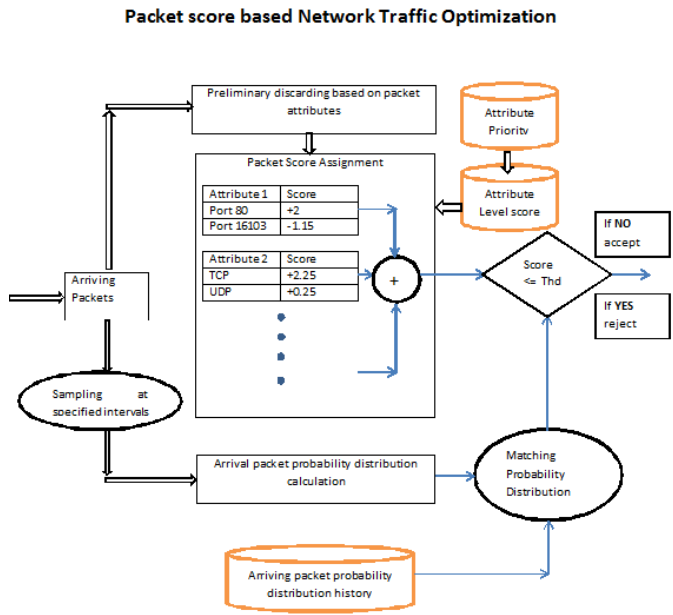

**Packet score based Network Traffic Optimization**

Fig.2.Packet score assignment and Traffic optimization

## VI. CONCLUSIONS

We have outlined the process the packet Score scheme used to defend against DDoS attacks. The key concept in packet Score is the Conditional Legitimate Probability (CLP) and probability distribution of arriving packets produced by comparison of legitimate traffic and attack traffic characteristics, which indicates the likelihood of legitimacy of a packet. As a result, packets following a legitimate traffic profile have higher scores, while attack packets have lower scores. This scheme can tackle never-before-seen DDoS attack types by providing a statistics-based adaptive differentiation between attack and legitimate packets to drive selective packet discarding and overload control at high- speed. Thus, packet Score is capable of blocking virtually all kinds of attacks as long as the attackers do not precisely mimic the sites' traffic characteristics. We have studied the performance and design tradeoffs of the proposed packet scoring scheme in the context of a stand-alone implementation. The newer simulation results in this paper are consistent with our previous research [19]. By exploiting the measurement/scorebook generation process, an attacker may try to mislead packet Score by changing the attack types and/or intensities. We can easily overcome such an attempt by using a smaller measurement period to track the attack traffic pattern more closely.

We are currently investigating the generalized implementation of packet Score for core networks. Packet Score is suitable for the operation at the core network at high speed, and we are working on an enhanced scheme for core network operation in a distributed manner. In particular, we plan to investigate the effects of update and feedback delays in a distributed implementation, and implement the scheme in hardware using network processors. Second, Packet Score is designed to work best for a large volume attack and it does not work well with low-volume attacks. We intend to explore and improve Packet Score performance in the presence of such attack types, e.g., bandwidth soaking attacks described in [31] or low-rate attacks [15]. Finally, a thorough investigation on the stability of traffic characteristics shall be performed as mentioned in Section 3